\documentclass[11pt,letterpaper]{article}

\usepackage[margin=1in]{geometry}
\usepackage{amsmath,amssymb,amsfonts,mathtools}
\usepackage{graphicx}
\usepackage{xcolor}
\usepackage{enumitem}
\usepackage{booktabs}
\usepackage{float}
\usepackage{setspace}

\usepackage{titlesec}
\usepackage{siunitx}
\usepackage{authblk}
\usepackage{csquotes}
\usepackage{fancyhdr}
\usepackage{subcaption}
\usepackage{caption}
\usepackage{multicol}
\usepackage{placeins}
\usepackage{epstopdf}
\graphicspath{{figures/}}
\DeclareGraphicsExtensions{.pdf,.png,.jpg}

\usepackage{hyperref}
\definecolor{linkcolor}{RGB}{0,60,120}

\usepackage[final,tracking=true,expansion=true,protrusion=true,nopatch=footnote]{microtype}

\usepackage[backend=biber,style=numeric,sorting=nyt,maxnames=6]{biblatex}
\AtEveryBibitem{
  \clearfield{urldate}
  \clearfield{note}
  \ifentrytype{online}
    {}
    {\iffieldundef{doi}{}{\clearfield{url}}}
}

\DeclareFieldFormat{eprint:arXiv}{
  arXiv\addcolon\space
  \ifhyperref
    {\href{https://arxiv.org/abs/#1}{#1}}
    {#1}
  \iffieldundef{eprintclass}
    {}
    {\addspace\mkbibbrackets{\strfield{eprintclass}}}
}
\DeclareFieldAlias{eprint:arxiv}{eprint:arXiv}

\AtEveryBibitem{
  \ifboolexpr{ test {\iffieldundef{eprint}} }
    {}
    { \clearfield{url} }
}

\setlist[itemize]{itemsep=2pt,topsep=2pt,leftmargin=1.5em}
\setlist[enumerate]{itemsep=2pt,topsep=2pt,leftmargin=1.5em}

\titleformat{\section}{\large\bfseries}{\thesection.}{0.5em}{}
\titleformat{\subsection}{\normalsize\bfseries}{\thesubsection}{0.5em}{}
\titleformat{\subsubsection}{\normalsize\itshape}{\thesubsubsection}{0.5em}{}

\hypersetup{
  colorlinks=true,
  urlcolor=blue,
  citecolor=black,
  linkcolor=black,
  pdfauthor={Aaryesh Deshpande},
  pdftitle={Learning Biomolecular Motion: The Physics-Informed Machine Learning Paradigm},
  pdfkeywords={physics-informed machine learning, biomolecular dynamics, PINN, differentiable simulation}
}

\newcommand{\keywords}[1]{\vspace{0.5em}\noindent\normalsize\textbf{Keywords:} #1\vspace{0em}}

\title{\textbf{Learning Biomolecular Motion:\\ The Physics-Informed Machine Learning Paradigm}}
\author[1]{Aaryesh Deshpande}
\affil[1]{School of Biological Sciences, Georgia Institute of Technology, Atlanta, GA \\
\texttt{adeshpande334@gatech.edu}}
\date{}

\begin{document}
\maketitle
\FloatBarrier 
\begin{abstract}
\noindent
The convergence of statistical learning and molecular physics is transforming our approach to modeling biomolecular systems. Physics-informed machine learning (PIML) offers a systematic framework that integrates data-driven inference with physical constraints, resulting in models that are accurate, mechanistic, generalizable, and able to extrapolate beyond observed domains. This review surveys recent advances in physics-informed neural networks and operator learning, differentiable molecular simulation, and hybrid physics-ML potentials, with emphasis on long-timescale kinetics, rare events, and free-energy estimation. We frame these approaches as solutions to the “biomolecular closure problem,” recovering unresolved interactions beyond classical force fields while preserving thermodynamic consistency and mechanistic interpretability. We examine theoretical foundations, tools and frameworks, computational trade-offs, and unresolved issues, including model expressiveness and stability. We outline prospective research avenues at the intersection of machine learning, statistical physics, and computational chemistry, contending that future advancements will depend on mechanistic inductive biases, and integrated differentiable physical learning frameworks for biomolecular simulation and discovery.
\end{abstract}


\section{Background and Scope}
\label{sec:background}

Molecular systems reside within highly complex, high-dimensional energy landscapes whose dynamics result from the coupled motion of thousands of atoms interacting via multiscale forces \cite{Hollingsworth2018-rs}. Classical molecular dynamics (MD) simulations, albeit based on first-principles physics, frequently experience restricted sampling efficiency, inadequate physical accuracy, and substantial computational expense \cite{Bernardi2015-js}. The challenge of connecting femtosecond bond vibrations to millisecond conformational shifts renders it practically unfeasible to thoroughly investigate the pertinent configurational space or to encapsulate emergent collective behavior \cite{Godwin2016-zq}. These inherent limitations stemming from the stiffness of atomic timescales, the curse of dimensionality, and the approximations in empirical potentials underscore the need for fast models that can infer long-timescale dynamics while retaining physical interpretability. 

Physics-informed machine learning (PIML) aims to overcome these obstacles by integrating \emph{physical principles} such as conservation laws, thermodynamic consistency, symmetries, and differential equation constraints into adaptable learning frameworks. PIML frameworks, in contrast to purely data-driven models, use inductive biases from established physical equations, such as laws of motion or free-energy correlations, to improve learning, strengthen generalization, and ensure interpretability \cite{Karniadakis2021-hr, thuerey2021pbdl}. 

\begin{figure}[H]
  \centering
     \includegraphics[width=0.75\textwidth]{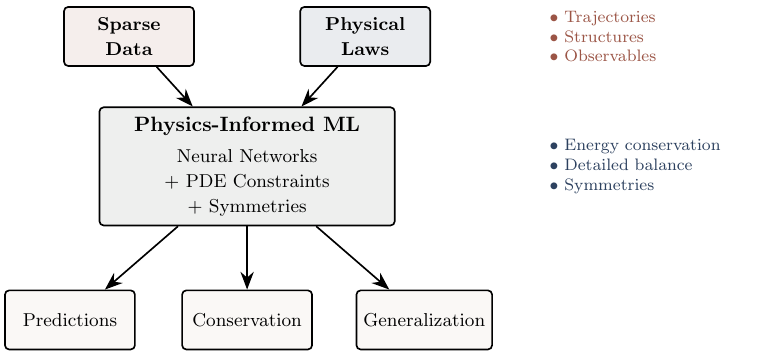}
  \caption{\textbf{The physics-informed machine learning paradigm}: Sparse observational data and governing physical laws are fused in neural models that encode conservation, symmetries, and differential constraints, yielding predictions that are accurate, stable, and generalizable.}
  \label{fig:piml_overview}
\end{figure}

The foundational principles of physics-informed learning can be traced to the early integration of statistical physics with neural computing, where energy-based formulations first linked physical models with learning dynamics. Early designs of energy-based neural networks, most notably the Hopfield network \cite{Hopfield1982-nw} and the Boltzmann machine \cite{hinton1985boltzmann} drew direct inspiration from spin-glass models in statistical physics \cite{Amit1985-ii}. These systems introduced the notion of a learnable energy landscape, in which the dynamics of neural states mirrored thermodynamic relaxation toward stable minima. By casting computation and memory as processes of energy minimization, they established one of the first formal bridges between physical law and learning dynamics. This perspective has since evolved into a broader unifying framework: as highlighted by Martin \emph{et al.} (2024), contemporary architectures such as Potts models, Boltzmann machines, and transformers~\cite{Vaswani2017-cu} share a common energy representation that links the principles of statistical mechanics with modern deep learning \cite{Martin2024-go}. These models reframed learning as an energy-minimization process analogous to the relaxation of a physical system toward equilibrium. The attractor states of Hopfield networks mirror metastable basins on molecular free-energy landscapes, while the Boltzmann distribution in probabilistic models parallels canonical ensembles in thermodynamics. This conceptual continuity directly influenced biomolecular modeling, where deriving energy functions from data aligns with the goal of approximating potential energy surfaces and conformational distributions \cite{Noe2019-dc,leach2013molecular}. Potts models extended these principles to biological sequences, revealing coevolutionary residue couplings and establishing the statistical basis for contact prediction and structural inference \cite{Morcos2011-ks, Marks2011-ad}. The same mathematical formalism now underlies energy-based transformers and equivariant graph neural networks that encode long-range couplings, symmetry constraints, and free-energy gradients in high-dimensional molecular systems \cite{Batzner2022-eh, Hu2021-kf, Martin2024-go}.

Over time, this lineage converged towards learning molecular physics directly from data under explicit physical constraints. Classical molecular dynamics, though foundational, relied on fixed empirical force fields such as AMBER and CHARMM \cite{Case2005-xg, Brooks2009-ug} with limited transferability and sampling efficiency. Machine learning-based potentials tried to address these shortcomings by learning the mapping between atomic environments and potential energies directly from high-level quantum data. The Behler-Parrinello neural network potential \cite{Behler2007-se} first demonstrated that atomic energy contributions could be learned through symmetry-aware descriptors, while Deep Potential Molecular Dynamics \cite{Zhang2018-zv} extended this principle to scalable many-body systems, achieving \emph{ab initio} accuracy at classical MD cost. To further ensure that learned interactions obeyed fundamental physical symmetries rather than relying on handcrafted features, subsequent models introduced explicit geometric equivariance. Building upon these foundations, E(3)-equivariant graph neural networks \cite{Batzner2022-eh, Hu2021-kf} implemented the encoding of atomic environments as geometric graphs that preserved rotational and translational invariance, enabling accurate prediction of molecular energies and forces while maintaining physical consistency across diverse structures.

The effort to model physical law directly into learning frameworks led to the development of Physics-Informed Neural Networks (PINNs), introduced by Raissi, Perdikaris, and Karniadakis \cite{Raissi2019-rz} and later unified under the broader PIML framework \cite{Karniadakis2021-hr}. PINNs formalized the idea of constraining neural models with governing differential equations, enforcing conservation laws, boundary conditions, and thermodynamic consistency during training rather than validating them post hoc. Building on this theoretical basis, differentiable simulation frameworks such as JAX-MD \cite{schoenholz2020jaxmd} and TorchMD-Net 2.0 \cite{thurlemann2024torchmdnet} operationalized these principles by coupling neural potentials with end-to-end differentiable molecular dynamics engines for the direct refinement of force fields and physical parameters under explicit dynamical constraints.

This review focuses on \emph{physics-informed machine learning} approaches that
\begin{enumerate}[label=(\roman*)]
\item \textbf{Embed physical laws explicitly} through constraints, differentiable operators, or inductive symmetries rather than using physics merely as inspiration;
\item \textbf{Model dynamical phenomena} like time evolution, free-energy reconstruction, kinetics, and reaction pathways emphasizing trajectories and transitions; and
\item \textbf{Evaluate performance under physical criteria}, including energy conservation, detailed balance, and thermodynamic consistency.
\end{enumerate}

Topics such as static structure prediction or sequence-based protein design are discussed only insofar as they inform or constrain dynamic modeling. 

A unifying perspective underpinning this review is that the biomolecular analogue of the \emph{closure problem} lies in representing unresolved interactions that classical potentials neglect electronic polarization, many-body effects, solvent coupling, and collective-variable projections. Conventional coarse-grained and empirical force fields approximate these contributions through fitted parameters, which often limits transferability and physical consistency across thermodynamic conditions. Hybrid physics-ML frameworks address this gap by acting as \emph{molecular closure models} that learn corrective forces, free-energy terms, or latent couplings that reintroduce missing physics while preserving statistical-mechanical stringency. From PINN-regularized molecular dynamics to neural operators and differentiable simulation engines, these methods extend the fidelity of molecular modeling while maintaining interpretability and thermodynamic coherence. Their objective is not only to reproduce observables but also to enforce \emph{mechanistic consistency}, conservation of momentum and energy, stability of trajectories, and extensivity of free-energy surfaces across systems and state points.

\section{Physical Foundations}
\label{sec:foundations}

\subsection{Thermodynamics and Statistical Mechanics}
\label{subsec:thermodynamics}

Biomolecular motion is governed by the principles of thermodynamics and statistical mechanics. Each molecular configuration $\mathbf{x}$ lies on a high-dimensional potential energy surface $E(\mathbf{x})$, where equilibrium probabilities follow the Boltzmann distribution.
\begin{equation}
p(\mathbf{x}) \propto \exp[-E(\mathbf{x})/(k_B T)].
\end{equation}
Low-energy conformations, such as folded proteins or ligand-bound complexes, are exponentially favored, forming an equilibrium ensemble defined by the partition function
\begin{equation}Z = \int e^{-E(\mathbf{x})/k_B T}\,d\mathbf{x}\end{equation} 
The corresponding free energy, $F = -k_B T \ln Z$, quantifies the relative stability and accessibility of molecular states.

From this perspective, biomolecular dynamics can be viewed as stochastic diffusion over the energy landscape; molecules fluctuate within local minima and occasionally cross barriers into new conformations. These rare transitions underpin folding, binding, and allosteric regulation, yet are notoriously difficult to capture via brute-force molecular dynamics (MD) simulations \cite{Hartmann2013-sn}. The underlying stochastic process is often modeled by the overdamped Langevin equation,
\begin{equation}
\frac{d\mathbf{x}}{dt} = -\nabla V(\mathbf{x}) + \boldsymbol{\xi}(t),
\end{equation}
where $\boldsymbol{\xi}(t)$ represents Gaussian thermal noise. The corresponding Fokker-Planck equation (utilized to represent the temporal evolution of the probability density function for systems influenced by stochastic variations) for the probability density $\rho(\mathbf{x},t)$ ensures that $\rho(\mathbf{x},t)\!\to\!p(\mathbf{x})$ as $t\!\to\!\infty$, enforcing detailed balance and thermodynamic consistency \cite{Pawula1967-zb}.

These principles define the physical constraints that any machine learning model of molecular motion must respect. A physically credible model should reproduce Boltzmann-weighted ensembles, conserve energy, and maintain equilibrium state populations. Generative formulations such as Boltzmann Generators try to achieve this by embedding $\exp(-E/k_B T)$ directly within the training objective, enabling unbiased equilibrium sampling and free-energy estimation. Likewise, physics-informed representations of stochastic dynamics try to employ neural networks to approximate drift or diffusion operators within the Langevin \cite{jeong2025lpinn} or Fokker-Planck frameworks \cite{Hu2024-kz}, ensuring compliance with statistical mechanical laws.

\subsection{Stochastic Thermodynamics and Nonequilibrium Extensions}
\label{subsec:nonequ}
Since biomolecular machines often operate far from equilibrium, their dynamics are naturally described at the trajectory level by stochastic thermodynamics, which assigns work $W$, heat $Q$, and entropy production $\Sigma$ to individual stochastic paths. Fluctuation theorems provide exact constraints that any physically credible learning model should respect; the Jarzynski equality $\langle e^{-\beta W}\rangle = e^{-\beta \Delta F}$ connects nonequilibrium work to equilibrium free-energy differences \cite{Jarzynski1997-ug}, while Crooks’ theorem relates the probabilities of forward and reverse paths, $P_F(\Gamma)/P_R(\tilde{\Gamma})=e^{\beta(W-\Delta F)}$ \cite{Crooks1999-xv}. These identities furnish trainable consistency checks and priors for learned dynamics and free-energy estimators.

In physics-informed settings, SDE-PINNs  and Langevin-PINNs fit the drift $b_\theta(\mathbf{x})$ and diffusion $D_\theta(\mathbf{x})$ of overdamped or underdamped Langevin models while enforcing thermodynamic structure and detailed balance at equilibrium along with nonnegative entropy production for driven steady states (NESS) \cite{OLeary2022-ev, jeong2025lpinn}. Practically, this can be realized by augmenting residual losses with (i) reversibility or skew-symmetry penalties, (ii) KL terms that match path-measure ratios implied by Crooks/Jarzynski, and (iii) constraints that couple $b_\theta$ and $D_\theta$ to known temperature and friction. Conceptually, these constraints align time-dependent PINNs with modern generative flows for molecules, where learned scores or transports are regularized toward Boltzmann or tilted (driven) ensembles while preserving microscopic reversibility whenever appropriate \cite{Noe2019-dc,Seifert2012-nu}.

\subsection{Variational Principles, PDE Constraints, and Inductive Bias}
\label{subsec:variational}
A unifying theme across physical sciences is that the governing equations of motion and equilibrium emerge from variational principles. Newton’s equations arise from minimizing the classical action, equilibrium corresponds to minimizing the free energy, and stochastic trajectories in high-friction regimes follow the most probable path defined by the Onsager-Machlup action (a function that encapsulates the dynamics of a continuous stochastic process). These principles provide a natural source of inductive bias that the system’s dynamics or configurations are not arbitrary but extremize a function constrained by conservation and dissipation laws. Embedding such principles into learning architectures ensures that inferred dynamics remain consistent with the stationary and pathwise optimality conditions of physics \cite{Greydanus2019-hi}.

Most molecular-scale phenomena are described by differential equations, ordinary or partial, that encode conservation of mass, momentum, and energy, as well as thermodynamic driving forces. Incorporating these PDEs directly into the loss function constrains learning to remain physically admissible throughout space and time, not merely at sampled data points. In practice, these variational priors are enforced either as residual penalties (PINNs) or through differentiable simulators. Automatic differentiation provides exact derivatives for evaluating the loss, allowing the model to enforce differential constraints continuously during training \cite{Raissi2019-rz}. This has been applied to molecular diffusion, reaction-diffusion kinetics, and quantum dynamics, where explicit trajectories are expensive or incomplete \cite{Karniadakis2021-hr, Pateras2025-zl}.

Beyond soft PDE enforcement, inductive bias can be encoded in the architecture itself. Hamiltonian and Lagrangian neural networks learn energy or action functionals, $H_\theta(\mathbf{x},\mathbf{p})$ or $\mathcal{L}_\theta(\mathbf{x},\dot{\mathbf{x}})$ whose derivatives yield physically consistent equations of motion \cite{Greydanus2019-hi, Cranmer2020-ej}. Such models conserve total energy and momentum by construction, making them well suited for learning biomolecular force fields $E_\theta(\mathbf{x})$ that remain differentiable and conservative across conformational space.

Variational formulations also account for the statistical treatment of molecular kinetics. The variational approach for Markov processes (VAMP) \cite{Peitz2019-cl, Wu2017-wr} defines optimal collective variables as those maximizing a variational score that approximates the leading eigenfunctions of the transfer operator. VAMPnets \cite{Mardt2018-th} and their modern extensions \cite{Schreiner2023-fk} operationalize this principle through neural representations that learn slow collective coordinates while respecting detailed balance and eigenfunction structure. These operator-based inductive biases yield models that reproduce long-time dynamics faithfully, stabilize training, and generalize beyond observed ranges. In this sense, physics-informed learning can be viewed as a continuum of variational and operator-constrained inference where minimizing action, residual, or spectral error collectively defines a unified and physically grounded approach to learning biomolecular dynamics.

\subsection{Operator Learning and Closure Modeling}
\label{subsec:olcm}
Beyond solving a single PDE instance, many biomolecular tasks require learning a \emph{solution operator} $\mathcal{G}:\mathcal{A}\!\to\!\mathcal{B}$ that maps inputs such as potentials, boundary/initial conditions, or parameters to fields (e.g., densities, fluxes). Operator-learning architectures approximate $\mathcal{G}$ directly; examples include DeepONet, which composes branch and trunk nets to achieve universal approximation of nonlinear operators \cite{Lu2021-qs}, while the Fourier Neural Operator (FNO) learns resolution-invariant mappings via spectral transforms \cite{Li2020-ia}. For MD-adjacent physics (Poisson-Boltzmann, Smoluchowski, Fokker-Planck), such operators deliver fast surrogates across families of systems rather than one-off solves, complementing PINNs that target a single instance \cite{Kadupitiya2020-gu, Schreiner2023-fk}. Mainly, operator learning provides a natural formalism for \emph{closure}, i.e., learning corrective maps that transfer between fine-grained and coarse-grained descriptions while preserving conservation laws, symmetry, and detailed balance. Embedding inductive biases-E(3) symmetry, conservative drifts, and fluctuation dissipation, turns these operators into \emph{molecular closure models} that recover missing many-body physics with thermodynamic consistency, setting up section~\ref{sec:frameworks} on differentiable hybrids and neural operators.

\section{Frameworks for Dynamics}\label{sec:frameworks}

\begin{figure}[H]
  \centering
     \includegraphics[width=0.85\textwidth]{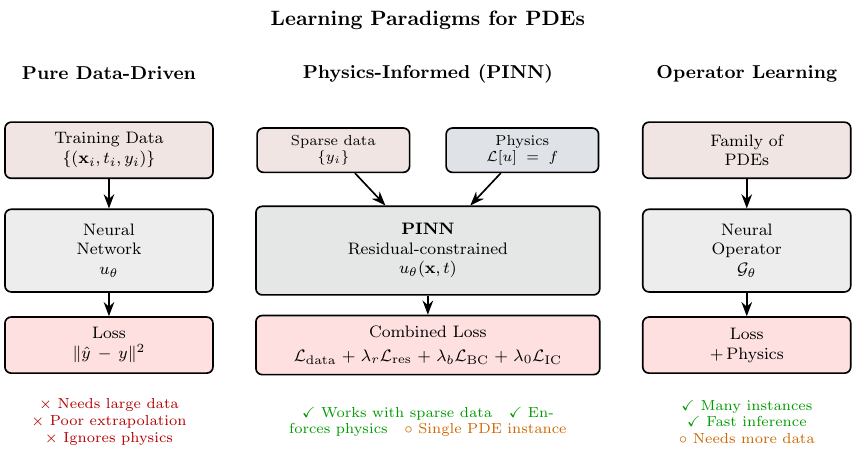}
  \caption{\textbf{Learning paradigms for PDE-governed systems:} Pure data-driven models fit observations but ignore physics; \emph{PINNs} add residual, boundary, and initial-condition losses to enforce the governing operator; \emph{operator learning} amortizes solutions across families of PDE instances for fast inference.}
  \label{fig:pde_paradigms}
\end{figure}

\subsection{Physics-Informed Neural Networks and Operator Learning}\label{subsec:pinns}
\noindent
\textbf{From variational priors to actionable constraints:}

Section~\ref{subsec:variational} established that biomolecular motion arises from principles of variational optimality, 
where trajectories and ensembles extremize actions and free energies subject to conservation and dissipation laws.
Physics-informed neural networks (PINNs) operationalize this principle by \emph{turning the governing operator itself into supervision}. Instead of training only against data, a PINN penalizes violations of the PDE, boundary conditions, and initial data, thereby steering learning toward the physically admissible manifold \cite{Karniadakis2021-hr}.

\textbf{Formulation:} Let $u_\theta(\mathbf{x},t)$ denote a neural approximation to a physical field (e.g., probability density, potential, or committor function).  
For a governing PDE
\begin{equation} \label{pinn_loss}
\mathcal{L}[u](\mathbf{x},t;\boldsymbol{\mu}) = f(\mathbf{x},t), \qquad (\mathbf{x},t)\in\Omega\times[0,T],
\end{equation}
with parameters $\boldsymbol{\mu}$ (e.g., temperature, friction), boundary operator $\mathcal{B}$ on $\partial\Omega$, and initial data $u_0$, a standard Physics-Informed Neural Network (PINN) minimizes the composite loss

\begin{equation}
\label{pinn_full}
\begin{aligned}
\mathcal{J}(\theta) 
&= \underbrace{\frac{1}{N_d}\sum_{i=1}^{N_d}\big\|u_\theta(\mathbf{x}_i,t_i)-y_i\big\|^2}_{\text{data loss}}
+ \lambda_r\,\underbrace{\frac{1}{N_r}\sum_{j=1}^{N_r}\big\|\mathcal{L}[u_\theta](\tilde{\mathbf{x}}_j,\tilde{t}_j;\boldsymbol{\mu})
- f(\tilde{\mathbf{x}}_j,\tilde{t}_j)\big\|^2}_{\text{residual loss}} \\
&\quad + \lambda_b\,\frac{1}{N_b}\sum_{k=1}^{N_b}\big\|\mathcal{B}[u_\theta](\hat{\mathbf{x}}_k,\hat{t}_k)\big\|^2
+ \lambda_0\,\frac{1}{N_0}\sum_{\ell=1}^{N_0}\big\|u_\theta(\mathbf{x}_\ell,0)-u_0(\mathbf{x}_\ell)\big\|^2 .
\end{aligned}
\end{equation}

\noindent
\begin{itemize}[leftmargin=1.25em,itemsep=2pt]
    \item \textbf{Data loss:} Ensures that the network reproduces available observations $y_i$ at sampled space-time points $(\mathbf{x}_i,t_i)$.  
    These may correspond to measured potentials, densities, or trajectory snapshots.
    \item \textbf{Residual loss:} Enforces the governing physics by penalizing deviations from the differential operator $\mathcal{L}[u_\theta]=f$.  
    This term acts as a continuous physics constraint that guides the network toward physically admissible solutions even where data are scarce.
    \item \textbf{Boundary loss:} Applies boundary conditions (e.g., fixed potential, reflecting or absorbing walls) to ensure correct behavior on $\partial\Omega$.  
    In molecular systems, this may correspond to confined domains or solvent interfaces.
    \item \textbf{Initial-condition loss:} Ensures consistency with prescribed initial states $u_0$ (e.g., starting distribution or structure).  
    This is particularly relevant for time-dependent MD-inspired formulations.
\end{itemize}

\noindent
Each coefficient $\lambda_{\{r,b,0\}}$ balances the influence of physical constraints against data fidelity, 
weighting the PDE residual, boundary, and initial-condition terms in the total loss.  Proper tuning of these hyperparameters is essential to avoid overemphasizing data or physics; adaptive weighting strategies are often used to maintain stability and physical consistency during training. Automatic differentiation provides exact spatial and temporal derivatives of $u_\theta$, enabling the loss to be computed directly from the network without discretization.

In this sense, PINNs belong to the broader class of \emph{physics-based deep learning} frameworks \cite{thuerey2021pbdl}, where model equations appear as differentiable constraints within the training objective. While differentiable-physics systems embed numerical solvers into the computation graph, PINNs treat the governing PDE itself as the residual term of the loss \cite{Raissi2019-rz,Karniadakis2021-hr}. Neural-operator methods such as DeepONet \cite{Lu2021-qs} and the Fourier Neural Operator (FNO) \cite{Li2020-ia} generalize this idea to learn the mapping from PDE inputs to solutions, enabling fast surrogate modeling across parameter families.

\textbf{Why this helps (ML intuition):}
Purely supervised fits can average over multi-modal solutions, drift off the conservation surface, and degrade over long rollouts.  
The residual term converts every space-time point into a weak label tied to the structural priors (invariance, locality, smoothness) inherent in $\mathcal{L}$.  
This (i) raises effective data density, (ii) curbs overfitting, and (iii) biases $u_\theta$ toward the PDE manifold, often yielding markedly better stability, detailed-balance consistency at equilibrium, and controlled entropy production in driven regimes (see \S.~\ref{subsec:nonequ}).

In a general biomolecular dynamics setting, $\mathcal{L}$ typically could encode overdamped/underdamped Langevin-Fokker-Planck dynamics (see \S~\ref{subsec:thermodynamics}), reaction-diffusion kinetics, or the backward Kolmogorov/committor equations. Enforcing these constraints stabilizes learning and promotes out-of-sample generalization along physically plausible directions.
 
PINNs excel when a \emph{single} PDE instance must be solved under scarce or partial observations or when hard-to-observe quantities (fluxes, free-energy gradients) are implicitly constrained by the operator.  
They become less attractive when many solvers are required across varying inputs (potentials, boundary data, temperatures), where training a new PINN per instance is inefficient. 
Many biomolecular workflows require solving whole \emph{families} of PDEs as conditions change, e.g., electrostatics with different boundary charges or ligand poses, Fokker-Planck dynamics across temperatures, solvent parameters, etc.
Neural operator learning addresses this by approximating the \emph{solution operator.}

\paragraph{When to use what?}
\begin{itemize}[leftmargin=1.25em,itemsep=2pt]
  \item \emph{Use PINNs} for \textbf{single, data-scarce inverse or forward problems} where physics must be guaranteed (e.g., committor fields, reaction-diffusion with sparse observables, or drift/diffusion identification subject to thermodynamic constraints).
  \item \emph{Use operator learning} when you need \textbf{many solvers} across varying inputs (geometry, BC/IC, $\boldsymbol{\mu}$).  
  In that case, we regularize with physics-residual losses, spectral penalties, or equilibrium/NESS constraints to retain fidelity, which is an idea central to modern surrogate modeling in biomolecular physics.
\end{itemize}

\subsection{Ecosystem and Tooling: Major PIML Libraries}\label{subsec:piml-libraries}
PINNs and operator-learning ideas are now supported by a mature, multi-language ecosystem. 
Table~\ref{tab:piml_software} summarizes widely used open-source libraries that provide 
high-level APIs for PINNs, neural operators, and related physics-based deep learning workflows. 
These tools differ in backends (TensorFlow, PyTorch, JAX, Julia), how directly they expose PDE 
residuals and boundary/initial conditions, and whether they function as stand-alone solvers or 
“wrappers” around deep-learning stacks. In section \ref{sec:difftools} we complement this view with 
differentiable \emph{molecular dynamics} engines (TorchMD, JAX-MD, TorchSim), which target 
trajectory-level optimization rather than PDE residual solves.

\begin{table}[H]
\centering
\small
\caption{Major software libraries for physics-informed ML (PIML). 
“Solver” indicates a library designed to specify PDE/ODE residuals, IC/BCs, and train end-to-end; 
“Wrapper” indicates a higher-level API that streamlines PIML model definition on top of a DL backend.}
\label{tab:piml_software}
\renewcommand{\arraystretch}{1.0}
\begin{tabular}{@{}llll@{}}
\toprule
\textbf{Software} & \textbf{Language} & \textbf{Backend} & \textbf{Usage} \\
\midrule
DeepXDE\cite{Lu2021-um}       & Python & TensorFlow, PyTorch, JAX, PaddlePaddle & Solver \\
NVIDIA PhysicsNeMo* \cite{Hennigh2021-en} & Python & PyTorch                                 & Solver \\
SciANN\cite{Haghighat2021-op}         & Python & TensorFlow, Keras                       & Wrapper \\
TensorDiffEq\cite{McClenny2021-zd}  & Python & TensorFlow                              & Solver \\
IDRLnet\cite{Peng2021-ty}        & Python & PyTorch                                 & Solver \\
NeuralPDE\cite{Zubov2021-ea}      & Julia  & Julia                                   & Solver \\
PND\cite{Razakh2021-ju}            & C++    & PyTorch                                 & Solver \\
NeuroDiffEq\cite{Chen2020-zd}    & Python & PyTorch                                 & Solver \\
PyDEns\cite{Koryagin2019-wh}         & Python & TensorFlow                              & Solver \\
ADCME\cite{Xu2020-et}          & Julia  & Julia, TensorFlow                       & Wrapper \\
Nangs\cite{Uriarte2024-lf}          & Python & PyTorch                                 & Solver \\
Elvet\cite{Araz2021-bh}          & Python & TensorFlow                              & Solver \\
\bottomrule
\end{tabular}

\vspace{0.4em}
\footnotesize\emph{Note:} Table adapted from the survey in Industrial \& Engineering Chemistry Research (ACS) \cite{Wu2023-rl}. 

*\small{Nvidia PhysicsNemo was formerly called Modulus/SimNet \cite{physicsnemo-docs}.}
\end{table}

While most PIML libraries (Table~\ref{tab:piml_software}) are frameworks focused on residual-driven PDE learning, emerging hybrid engines extend these ideas to physically constrained simulation. 
\textbf{PND} (Physics-informed Neural Dynamics) \cite{Razakh2021-ju} exemplifies the high-performance end of this spectrum; it embeds PINN solvers directly into a parallel molecular dynamics (MD) engine. 
This design enables simultaneous enforcement of conservation laws and least-action principles during atomistic simulations, 
providing training of physics-informed potentials and closure terms within traditional MD workflows. 

At the opposite end, \textbf{TorchSim} \cite{Cohen2025-sf} (see \S~\ref{sec:difftools}) represents a fully differentiable, GPU-native platform written 
entirely in PyTorch. It integrates automatic differentiation through all simulation components, energies, forces, thermostats, 
and integrators, enabling end-to-end gradient propagation from trajectory- or experiment-level objectives. 

While these libraries target \emph{residual-based} learning (PINNs, neural operators), 
trajectory-centric learning is enabled by differentiable MD engines (JAX-MD, TorchMD-Net~2.0, TorchSim; 
see \S~\ref{sec:difftools}), which expose integrators and thermostats to autodiff and support end-to-end 
training from observable or experiment-level objectives.

\subsection{Differentiable Simulation and Hybrid Physics-ML}\label{sec:difftools}

A complementary paradigm to physics-informed supervision is to \emph{differentiate through the simulator itself}.  
In \textbf{differentiable physics} (DP) frameworks \cite{thuerey2021pbdl,schoenholz2020jaxmd,thurlemann2024torchmdnet}, the discrete equations of motion are embedded directly within the computational graph, allowing gradients to flow through numerical integrators.  
Rather than enforcing physical laws as external losses, DP composes neural components (e.g., potentials, forces, or control fields) with a differentiable time-stepping solver and optimizes them end-to-end using backpropagation.

\paragraph{Formulation:}
Let $\Phi_{\Delta t}$ denote a (possibly stochastic) molecular dynamics (MD) update operator that advances the system state 
$\mathbf{z}_t = (\mathbf{x}_t,\mathbf{v}_t)$ comprising positions and velocities by one integration step under learnable parameters $\boldsymbol{\phi}$ (such as neural potential weights, friction coefficients, or thermostat controls):
\begin{equation}
\mathbf{z}_{t+1} = \Phi_{\Delta t}(\mathbf{z}_t;\boldsymbol{\phi}).
\end{equation}
Unrolling $T$ integration steps defines a differentiable simulation trajectory 
$\{\mathbf{z}_0,\mathbf{z}_1,\ldots,\mathbf{z}_T\}$ with a task-specific objective
\begin{equation}
\mathcal{L}(\boldsymbol{\phi}) 
= \sum_{t=0}^{T} \ell\big(\mathbf{z}_t;\,\text{targets}\big),
\end{equation}
where $\ell$ measures discrepancy between simulated and reference quantities (e.g., forces, radial distributions, free energies, or kinetic observables).  
Gradients $\nabla_{\boldsymbol{\phi}}\mathcal{L}$ are obtained by backpropagation through time (BPTT), with memory requirements managed via \emph{checkpointing} or \emph{implicit differentiation}.  
For stochastic integrators (e.g., Langevin dynamics), differentiable noise reparameterizations or likelihood-ratio estimators provide unbiased, low-variance gradient estimates.

This paradigm enables:
\begin{itemize}[leftmargin=1.25em,itemsep=2pt]
  \item \textbf{Learning potentials or force fields} directly from trajectories or experimental observables, ensuring microscopic consistency;
  \item \textbf{Calibrating thermostats and friction} to reproduce desired kinetic or diffusive behaviors;
  \item \textbf{Optimizing biasing and control protocols} for enhanced or adaptive sampling;
  \item \textbf{Solving inverse problems amortized over simulations}, such as coarse-grained (CG) parameterization, while retaining physical guarantees.
\end{itemize}
By coupling neural approximators with differentiable integrators, DP unifies simulation and learning into a single optimization loop, bridging the gap between mechanistic models and trainable surrogates.

\textbf{Differentiable MD Engines:}
The differentiable-physics paradigm is now embodied in molecular simulation frameworks that expose integrators, forces, and thermodynamic operators to automatic differentiation as discussed briefly in section~\ref{subsec:piml-libraries}. Some of these engines are listed below:

\begin{figure}[H]
  \centering
   \includegraphics[width=0.75\textwidth]{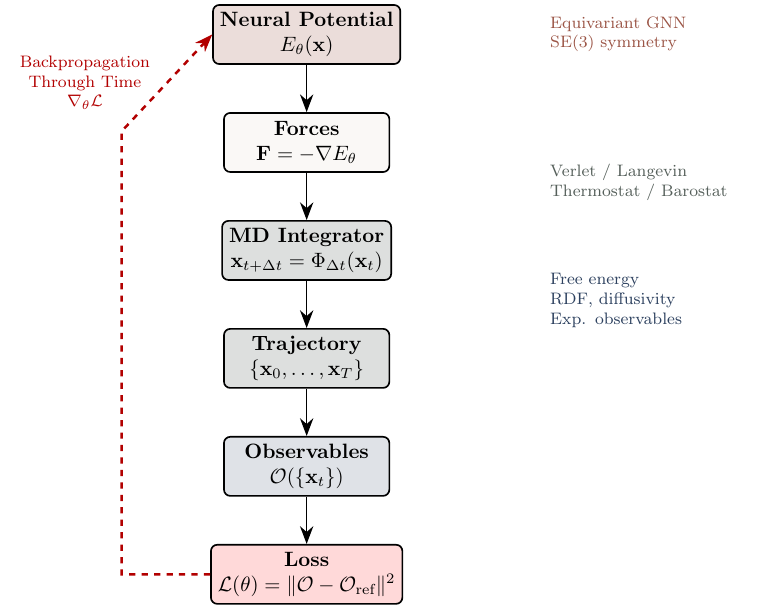} 
  \caption{\textbf{An ideal End-to-end differentiable simulation:} A learnable potential produces forces that drive a differentiable MD integrator; trajectories yield observables compared to references, and gradients are backpropagated through time to update model parameters, enabling calibration to ensemble and kinetic targets.}
  \label{fig:diff_eng}
\end{figure}

\begin{itemize}[leftmargin=1.25em,itemsep=3pt]
  \item \textbf{JAX-MD} \cite{schoenholz2020jaxmd} provides fully differentiable energies, neighbor lists, and integrators, allowing gradients to flow end-to-end from observables (e.g., free energies, diffusivities) to force-field parameters.
  \item \textbf{TorchMD / TorchMD-Net 2.0} \cite{Doerr2021-fy, thurlemann2024torchmdnet} couples equivariant GNN potentials with differentiable MD loops in PyTorch, supporting large-scale, multi-GPU backpropagation through trajectories.
  \item \textbf{DiffTaichi} and \textbf{PhiFlow} \cite{thuerey2021pbdl} extend these ideas to continuum and fluid systems, enabling differentiable PDE solvers and hybrid neural-physical models.
  \item \textbf{TorchSim} \cite{Cohen2025-sf} extends differentiable MD to large-scale, batched, GPU-native atomistic simulations implemented entirely in PyTorch. 
TorchSim introduces \emph{AutoBatching}, which automatically packs multiple heterogeneous systems into GPU memory for concurrent time integration, maximizing utilization across MLIPs such as MACE, MatterSim, PET-MAD, EGIP, and SevenNet.
It supports both deterministic (e.g., NVE/NVT/NPT) and stochastic (Langevin) integrators, provides differentiable access to energies, forces, and observables through PyTorch’s autograd, and allows end-to-end optimization of potentials from trajectory- or experiment-level objectives.
TorchSim thus merges the high-level usability of TorchMD with the scalability of traditional MD packages, achieving up to two orders of magnitude higher throughput on modern GPUs through batched simulations.
\end{itemize}

Rather than fitting to precomputed data, these engines differentiate through integrators themselves to learn dynamics consistent with physical laws.  
Emerging trend here is to integrate backends, such as differentiable modules in \textbf{LAMMPS}, \textbf{OpenMM}, and related engines, to bridge ML differential frameworks with established MD infrastructure for scalable, physics-constrained training.

\textbf{Integrated vs.\ Differentiable ML Potentials:}
Classical engines (\textbf{LAMMPS} \cite{Barnes2017-ev}, \textbf{OpenMM} \cite{Eastman2024-qz}) load \emph{pretrained} ML potentials (e.g., DeepMD-kit \cite{Wang2018-is}, NequIP \cite{Batzner2022-eh}, TorchANI \cite{Gao2020-vg}) and evaluate forces during MD; the solver itself is non-differentiable, so parameter learning occurs offline (\emph{inference mode}). By contrast, frameworks such as \textbf{JAX-MD} and \textbf{TorchMD-Net~2.0} expose forces, thermostats, and integrators to autodiff, enabling end-to-end optimization of potentials, CG closures, and control signals directly from trajectory-level losses.

\begin{table}[H]
\centering
\small
\caption{Comparison of differentiable and hybrid MD engines for physics-informed simulation. 
Green (\textcolor{green!60!black}{\checkmark}) indicates full support, yellow (\textcolor{orange!80!black}{\(\odot\)}) partial support, and red (\textcolor{red!70!black}{\(\times\)}) lack of support. 
TorchSim provides full differentiability, PyTorch-native batching, and integration with modern MLIPs.}
\label{tab:diffmd_engines}
\renewcommand{\arraystretch}{1.15}
\begin{tabular}{@{}lcccccc@{}}
\toprule
\textbf{Feature} & \textbf{OpenMM} & \textbf{LAMMPS} & \textbf{TorchMD} & \textbf{ASE} & \textbf{JAX-MD} & \textbf{TorchSim} \\ 
\midrule
Batching & \textcolor{red!70!black}{\(\times\)} & \textcolor{red!70!black}{\(\times\)} & \textcolor{orange!80!black}{\(\odot\)} & \textcolor{red!70!black}{\(\times\)} & \textcolor{orange!80!black}{\(\odot\)} & \textcolor{green!60!black}{\(\checkmark\)} \\ 
Diverse MLIPs & \textcolor{red!70!black}{\(\times\)} & \textcolor{red!70!black}{\(\times\)} & \textcolor{red!70!black}{\(\times\)} & \textcolor{green!60!black}{\(\checkmark\)} & \textcolor{red!70!black}{\(\times\)} & \textcolor{green!60!black}{\(\checkmark\)} \\ 
Differentiable & \textcolor{red!70!black}{\(\times\)} & \textcolor{red!70!black}{\(\times\)} & \textcolor{green!60!black}{\(\checkmark\)} & \textcolor{red!70!black}{\(\times\)} & \textcolor{green!60!black}{\(\checkmark\)} & \textcolor{green!60!black}{\(\checkmark\)} \\ 
Pure Python & \textcolor{red!70!black}{\(\times\)} & \textcolor{red!70!black}{\(\times\)} & \textcolor{green!60!black}{\(\checkmark\)} & \textcolor{green!60!black}{\(\checkmark\)} & \textcolor{green!60!black}{\(\checkmark\)} & \textcolor{green!60!black}{\(\checkmark\)} \\ 
GPU Dynamics & \textcolor{green!60!black}{\(\checkmark\)} & \textcolor{green!60!black}{\(\checkmark\)} & \textcolor{green!60!black}{\(\checkmark\)} & \textcolor{red!70!black}{\(\times\)} & \textcolor{green!60!black}{\(\checkmark\)} & \textcolor{green!60!black}{\(\checkmark\)} \\ 
Multi GPU & \textcolor{green!60!black}{\(\checkmark\)} & \textcolor{green!60!black}{\(\checkmark\)} & \textcolor{red!70!black}{\(\times\)} & \textcolor{red!70!black}{\(\times\)} & \textcolor{red!70!black}{\(\times\)} & \textcolor{orange!80!black}{\(\odot\)} \\ 
Integration w/ MLIPs & \textcolor{orange!80!black}{\(\odot\)} & \textcolor{orange!80!black}{\(\odot\)} & \textcolor{green!60!black}{\(\checkmark\)} & \textcolor{red!70!black}{\(\times\)} & \textcolor{green!60!black}{\(\checkmark\)} & \textcolor{green!60!black}{\(\checkmark\)} \\ 
Auto Memory Mgmt & \textcolor{red!70!black}{\(\times\)} & \textcolor{red!70!black}{\(\times\)} & \textcolor{red!70!black}{\(\times\)} & \textcolor{red!70!black}{\(\times\)} & \textcolor{orange!80!black}{\(\odot\)} & \textcolor{green!60!black}{\(\checkmark\)} \\ 
\bottomrule
\end{tabular}
\end{table}

\subsection{Hybrid Physics-ML taxonomies}\label{subsec:hybrid-physics}
Hybrid models integrate established physical laws with learnable \emph{closures} that account for unresolved interactions such as polarization, many-body coupling, and solvent effects.  
Rather than replacing physics outright, they introduce a flexible correction $E_\theta$ that complements a trusted baseline potential $E_{\mathrm{phys}}$, yielding a \emph{hybrid energy landscape}:
\begin{equation}
E_{\mathrm{hyb}}(\mathbf{x}) = E_{\mathrm{phys}}(\mathbf{x}) + E_\theta(\mathbf{x}).
\label{eq:hybrid-energy}
\end{equation}
The learned term $E_\theta$ adapts to missing physics while retaining the variational structure, ensuring that dynamics derived from $E_{\mathrm{hyb}}$ remain conservative and differentiable.

\textbf{Hybrid Langevin formulations:}
For molecular systems evolving under Langevin dynamics, the hybrid potential modifies the stochastic equations of motion as
\begin{equation}
m\ddot{\mathbf{x}} + \gamma\,\dot{\mathbf{x}} 
+ \nabla E_{\mathrm{phys}}(\mathbf{x}) 
+ \nabla E_\theta(\mathbf{x}) 
\;=\; \boldsymbol{\eta}(t),
\qquad 
\langle \eta_i(t)\eta_j(t')\rangle 
= 2k_BT\gamma\,\delta_{ij}\delta(t-t'),
\label{eq:underdamped-hybrid}
\end{equation}
where $\mathbf{x}$ are particle coordinates, $m$ the mass matrix, $\gamma$ the friction coefficient, and $\boldsymbol{\eta}(t)$ a Gaussian stochastic force satisfying the fluctuation-dissipation theorem (FDT).  
Equation~\eqref{eq:underdamped-hybrid} represents the \emph{underdamped hybrid Langevin equation}, ensuring that the equilibrium distribution $p(\mathbf{x})\!\propto\!\exp[-E_{\mathrm{hyb}}(\mathbf{x})/k_BT]$ is preserved if FDT holds.

In the high-friction (overdamped) regime, velocities equilibrate quickly, leading to a stochastic differential equation for positional diffusion:
\begin{equation}
\dot{\mathbf{x}} 
= -D(\mathbf{x})\,\nabla\!\Big(E_{\mathrm{phys}}(\mathbf{x})+E_\theta(\mathbf{x})\Big)
+ \sqrt{2D(\mathbf{x})}\,\boldsymbol{\xi}(t),
\qquad
\langle \xi_i(t)\xi_j(t')\rangle=\delta_{ij}\delta(t-t'),
\label{eq:overdamped-hybrid}
\end{equation}
where $D(\mathbf{x})$ is the position-dependent diffusion tensor linked to $\gamma$ via FDT ($D=k_BT/\gamma$ for constant friction)?  
Equations~\eqref{eq:underdamped-hybrid}-\eqref{eq:overdamped-hybrid} define a \textbf{thermodynamically consistent hybrid simulator} that is physical in its structure yet learnable in its unresolved components.

\begin{itemize}[leftmargin=1.25em,itemsep=3pt]
\item $\nabla E_{\mathrm{phys}}(\mathbf{x})$: Deterministic physical forces (e.g., bonded, van der Waals, electrostatic) from established force fields or coarse-grained models.
\item $\nabla E_\theta(\mathbf{x})$: Data-driven corrective forces that restore missing physical contributions learned via neural potentials or differentiable simulation frameworks.
\item $\gamma\,\dot{\mathbf{x}}$: Viscous damping or solvent drag that dissipates energy at a rate consistent with frictional coupling.
\item $\boldsymbol{\eta}(t)$ and $\boldsymbol{\xi}(t)$: Stochastic driving forces representing thermal fluctuations, constrained by FDT to maintain canonical equilibrium.
\item $D(\mathbf{x})$: Diffusion tensor, possibly learned as $D_\theta(\mathbf{x})$ in anisotropic or spatially heterogeneous environments.
\end{itemize}

This formalism connects seamlessly to biomolecular simulation practice, as  
the deterministic component ensures faithful force reproduction,  
the stochastic term captures solvent-mediated noise,  
and the learned closure encapsulates missing many-body or environmental effects.  
Differentiable MD engines (e.g., TorchMD, JAX-MD) implement these equations directly, enabling end-to-end learning of $E_\theta$ or $D_\theta$ through backpropagation from trajectory or observable-level losses while maintaining compliance with equilibrium thermodynamics. \textbf{What gets learned?} Hybrid frameworks differ primarily in the nature of the learned correction or closure taxonomy:
\begin{itemize}[leftmargin=1.25em,itemsep=2pt]
\item \textbf{Residual conservative forces} $\nabla E_\theta$: Small, symmetry-preserving corrections trained on quantum-mechanical energies, forces, or experimental observables; maintain conservative structure by construction.
\item \textbf{Coarse-grained (CG) closures}: Mappings from atomistic (AA) statistics to CG forces that satisfy detailed balance and FDT, enabling accurate transfer across thermodynamic states \cite{Wang2019-hi}.
\item \textbf{Generative priors}: Score-based or normalizing-flow models regularizing sampled ensembles toward Boltzmann while allowing controlled deviations for nonequilibrium or biased dynamics \cite{Noe2019-dc}.
\item \textbf{Dissipative structure}: Spatially varying $D_\theta(\mathbf{x})$ or friction fields that adapt dissipation rates while remaining thermodynamically admissible ($D_\theta = k_BT/\gamma_\theta$).
\end{itemize}

Hybrid training typically blends multi-scale criteria:
\begin{align*}
\mathcal{L} \;=\; 
&\underbrace{\alpha_E\,\|E_\theta+E_{\mathrm{phys}}-E^{\mathrm{ref}}\|_2^2 
+ \alpha_F\,\|F_\theta+F_{\mathrm{phys}}-F^{\mathrm{ref}}\|_2^2}_{\textit{microscopic (QM/AA) fits}}
\\
&+ \underbrace{\alpha_{\mathrm{ens}}\Big[\mathrm{KL}\!\big(p_\theta\|\;p_{\mathrm{ref}}\big) 
+ \lambda_{\Delta F}\,|\Delta F_\theta-\Delta F_{\mathrm{ref}}|\Big]}_{\textit{mesoscopic ensembles/free energies}}
\\
&+ \underbrace{\alpha_{\mathrm{dyn}}\Big[\mathrm{MSE}(\mathrm{VACF}),\,|\!D_\theta-D_{\mathrm{ref}}|,\,|\mathrm{MFPT}_\theta-\mathrm{ref}|\Big]}_{\textit{dynamics/kinetics}}
\\
&+ \underbrace{\alpha_{\mathrm{exp}}\Big[\mathrm{MSE}(\mathrm{SAXS}),\,\mathrm{MSE}(\mathrm{RDF}/S(q)),\,\mathrm{NMR(RDC/NOE)},\,\mathrm{cryo\text{-}EM}\Big]}_{\textit{experiment-in-the-loop}}.
\end{align*}
Here $p_\theta$ is the model ensemble and $\Delta F$ denotes thermodynamic differences estimated via reweighting or nonequilibrium work identities. 
This formulation mirrors the base physics-informed loss but extends it across scales, linking microscopic force and energy fidelity to mesoscopic ensemble accuracy, dynamical observables, and experimental constraints, thereby adding data-driven and dynamic-ground objectives within a single differentiable framework. Differentiating through these long rollouts is memory intensive and sensitive to stiffness. Effective recipes include: (i) short-horizon unrolls with \emph{multiple shooting}; (ii) checkpointing or implicit/adjoint gradients; (iii) \emph{stage-wise training} (force matching $\rightarrow$ trajectory-/observable-level fine-tuning); and (iv) \emph{hard structure} (equivariance, conservative heads, FDT links) rather than only soft penalties \cite{thuerey2021pbdl,Batzner2022-eh,Doerr2021-fy}.

\paragraph{Where hybrids shine:}
\begin{itemize}[leftmargin=1.25em,itemsep=2pt]
\item \textbf{Ab initio accuracy at classical cost:} DP/NequIP/DeepPotential-style energies reach QM-grade forces with MD efficiency \cite{Zhang2018-zv,Batzner2022-eh}.
\item \textbf{Fast, faithful CG:} Learned closures stabilize CG kinetics and transfer across thermodynamic states \cite{Husic2020-vw}.
\item \textbf{Rare events and control:} Differentiable biasing learns protocols that accelerate transitions while preserving reweightability; kinetic targets can be optimized directly.
\item \textbf{Experiment closure:} Backpropagating from SAXS/NMR/cryo-EM to physical parameters closes the loop between data and simulation \cite{He2020-ua}
\end{itemize}

\begin{figure}[H]
  \centering
   \includegraphics[width=0.75\textwidth]{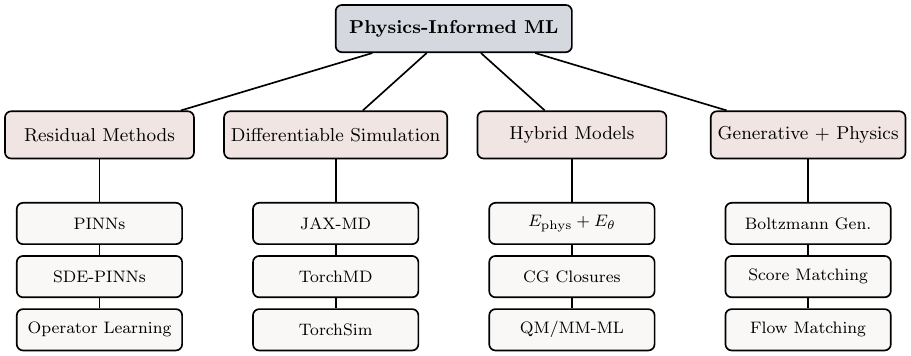} 
  \caption{\textbf{Taxonomy of physics-informed ML for biomolecular dynamics:}
  Four complementary strands; residual methods (PINNs, SDE-PINNs, neural operators), differentiable simulation (JAX-MD, TorchMD, TorchSim), hybrid physics–ML closures ($E_{\mathrm{phys}}{+}E_\theta$, CG, QM/MM-ML), and physics-aware generative models (Boltzmann generators, score/flow matching), span the space from equations to trajectories to ensembles.}
  \label{fig:tax}
\end{figure}

\section{Applications in Biomolecular Science}\label{sec:applications}

\subsection{Free-Energy Surface Learning}\label{subsec:FES}
The free-energy surface (FES) $F(s)=-k_\mathrm{B}T\ln p(s)$ over collective variables (CVs) $s$ encodes both thermodynamics and kinetics. 
Traditional enhanced-sampling methods rely on a few hand-crafted CVs; in contrast, physics-informed ML (PIML) \emph{learns} CVs and $F(s)$ jointly, guided by the principles mentioned in \S\ref{subsec:variational}-\ref{subsec:olcm}. 
The goal is not merely to interpolate energies but to respect equilibrium statistics, detailed balance, and (when needed) controlled departures in nonequilibrium settings.

\textbf{CV discovery as variational operator learning:}
Time-structure and transfer-operator perspectives motivate learning CVs that approximate slow eigenfunctions. 
VAMP/TICA objectives and their neural realizations (e.g., VAMPnets) maximize a variational score linked to the dominant spectrum of the transfer operator, providing data-driven, physically meaningful coordinates for metastable transitions \cite{Mardt2018-th, Wu2017-wr, Bonati2021-cx}. 
Equivariant graph encoders further enforce $E(3)$ symmetries and permutation invariances, yielding descriptor-free CVs consistent with molecular geometry \cite{Batzner2022-eh, Hu2021-kf}.

\textbf{Learning the FES with thermodynamic priors:}
Normalizing flows, score models, and Boltzmann Generators minimize divergence to the Boltzmann distribution, enabling direct equilibrium sampling and $\Delta F$ estimation from generated ensembles \cite{Noe2019-dc}. 
Hybrid closures (see \S\ref{subsec:hybrid-physics}) augment trusted baselines $E_{\mathrm{phys}}$ with learnable $E_\theta$ while maintaining a conservative structure so that $p(\mathbf{x})\!\propto\!\exp[-(E_{\mathrm{phys}}+E_\theta)/k_BT]$ remains valid. 
In differentiable MD (\S\ref{sec:difftools}), bias networks can be trained end-to-end by backpropagating discrepancies in $p(s)$ or reweighting estimators of $F(s)$.
Another approach for free-energy reconstruction employs Gaussian Process Regression (GPR), which incorporates the Jacobians of collective variables (CVs) to capture local geometric relationships in the reduced coordinate space \cite{Stecher2014-zf}. More recently, neural-network surrogate frameworks have been introduced that learn CVs directly from Cartesian coordinates and use automatic differentiation to compute the corresponding Jacobians. This eliminates the need for explicit analytical expressions and enables a fully differentiable, data-driven mapping between atomic configurations and free-energy landscapes \cite{Pornpatcharapong2025-zd}.

PIML turns FES learning into a \emph{variationally constrained and physically constrained} problem: CVs can be derived from spectral optimization (VAMP), GPR with Jacobians, or neural surrogates that learn CVs from Cartesian coordinates via automatic differentiation. The free energy $F(s)$ is refined through thermodynamic consistency (flows/Boltzmann generators) and differentiable physical losses, yielding bias-free, uncertainty-aware free-energy landscapes for kinetic modeling.

\subsection{Protein Folding and Dynamics}\label{sec:protein-dynamics}
Protein folding tests whether models capture \emph{both} equilibrium landscapes and long-time kinetics. 
The frameworks in section~\ref{sec:frameworks} naturally decompose the problem into (i) conservative energetics, (ii) slow-mode discovery, and (iii) trajectory-consistent learning.

\textbf{PINN-guided coarse-grained (CG) closures:}
Hybrid energies $E_{\mathrm{hyb}}=E_{\mathrm{phys}}+E_\theta$ (see \S\ref{subsec:hybrid-physics}) let $E_\theta$ encode missing many-body terms while preserving conservative structure. 
PINN-style residuals penalize violations of the governing dynamics, encouraging energy and momentum consistency during CG simulations. 
Stage-wise training and force matching $\rightarrow$ ensemble targets $\rightarrow$ kinetic observables, stabilizes learning, and improves transfer across temperatures and solvent conditions \cite{Doerr2021-fy, Batzner2022-eh}.

\textbf{Operator learning for kinetics:}
VAMPnets approximate leading eigenfunctions of the transfer operator, learning latent coordinates and coarse-grained Markov models directly from short MD \cite{Mardt2018-th}. This yields metastable state partitions, equilibrium populations, and implied timescales that extrapolate to millisecond dynamics when validated by Chapman-Kolmogorov tests \cite{Goswami2024-qh}.

\textbf{Generative ensemble models:}
Diffusion/flow-based generative models trained on simulation and structure repositories can emulate Boltzmann-consistent ensembles, filling gaps between metastable states and enabling rapid \emph{ensemble} predictions of observables. 
When coupled to differentiable simulators (see \S\ref{sec:difftools}), generative priors act as bias proposals that remain reweightable to equilibrium \cite{Noe2019-dc}. 
Recent large generative models for protein motion (e.g., diffusion models conditioned on sequence/structure) aim to blend pLM representations with physically grounded sampling to recover order parameters, relaxation spectra, and conformational variability \cite{Lewis2025-yb, Gelman2025-gv}.

Folding and dynamics modeling benefit from a clear division of labor, i.e., conservative, equivariant potentials that provide physically credible forces; operator-learning methods that identify slow collective coordinates and recover faithful kinetics; and differentiable simulators that couple these components in end-to-end optimization loops. In this way, physics-informed ML moves beyond static structure toward dynamic prediction, yielding ensembles calibrated directly to experimental observables.

\paragraph{Related perspective: PINNs in protein design:}
Complementary to dynamics, PINNs are being used for \emph{inverse} sequence-structure design as PDE-constrained optimization problems. 
Here, a PINN embeds a differentiable forward biophysical operator ranging from conservative energy models and elastic network surrogates to Poisson-Boltzmann or diffusion-type PDEs and searches over sequences/backbone tweaks under physics-based regularization (stability, foldability, interface energetics) and inductive biases (symmetries, conservation, detailed balance where relevant). 
The differentiable setup naturally handles boundary/initial conditions and enables gradient-guided exploration of large design spaces with interpretable constraints, while exposing characteristic training pathologies (operator stiffness, spectral bias, and the need to balance data vs.\ physics losses or to schedule them adaptively). 
Applications include stability-guided sequence search, epitope/paratope and interface design, and scaffold retargeting; promising hybrids couple PINNs with generative priors (flows, diffusion models, or pLMs) so that samples are steered toward physically admissible regions and subsequently refined by PDE residuals and ensemble-level criteria \cite{Omar2023-cq}.

\subsection{Ligand Binding and Catalysis}\label{sec:ligand-binding}
Binding and catalysis couple conformational dynamics with electronic rearrangements and solvent responses. 
PIML frameworks can be used for docking, free-energy estimation, and reaction-path analysis under explicit physical constraints (see \S\ref{subsec:thermodynamics}, \S\ref{subsec:nonequ}).

\textbf{Physics-aware docking and scoring:}
Diffusion-based pose generation (e.g., DiffDock-style denoising) can be regularized by differentiable physical layers, continuum electrostatics, desolvation, and steric penalties so that pose likelihoods correlate with, $\Delta G_\mathrm{bind}$ rather than purely geometric fit \cite{Corso2022-gh}. 
Normalizing-flow or score-heads trained with Boltzmann-consistency terms encourage physically plausible pose ensembles that remain reweightable to equilibrium.

\textbf{ML-QM/MM surrogates:}
Equivariant GNN potentials (NequIP/GemNet-class) trained on high-level QM data replace or assist the QM region in QM/MM, preserving rotational/permutation symmetries and achieving near-chemical accuracy while enabling reactive MD \cite{Batzner2022-eh, Gasteiger2021-tl}. 
Within the hybrid formulation (see \S\ref{subsec:hybrid-physics}), $E_\theta$ captures short-range electronic effects (polarization, charge transfer), while $E_{\mathrm{phys}}$ supplies long-range interactions and boundary conditions. 
This differentiable MD permits end-to-end calibration against reaction barriers, isotope effects, or catalytic turnover metrics.

\textbf{Learning reaction coordinates via committors:}
Reaction progress in complex enzymes is naturally parameterized by the committor $q(\mathbf{x})$, the probability of reaching products before reactants. 
Neural committor approximators $q_\theta(\mathbf{x})$ trained with operator or transition-path residuals learn nonlinear reaction coordinates; the hypersurface approximates the transition state and guides targeted sampling and PMF estimation \cite{Alagna2025-hs, Jung2023-wt}.

\section{Technical Limitations}\label{sec:limitations}
Despite rapid progress, physics-informed ML faces several fundamental limitations that must be acknowledged for responsible application. This section catalogs some failure modes and their possible underlying causes.

\subsection{The Extrapolation Problem:}

While PIML methods excel at interpolation within the convex hull of training data, extrapolation remains precarious. A force field trained on equilibrium structures may produce nonsensical energies for highly strained geometries encountered during rare events \cite{Kaser2023-mu}. For example, a protein-ligand binding model trained at 300~K may give qualitatively wrong predictions at 350~K or in different ionic strengths unless thermodynamic consistency is explicitly enforced. Similarly, models trained on folded conformations often fail catastrophically when applied to intrinsically disordered regions or unfolding intermediates. The root cause stems from the principle that neural networks are interpolators, not extrapolators \cite{Hooven2025-lt, Pun2019-yz}. Without explicit physical constraints, they learn spurious correlations specific to the training distribution. Energy functions may lack proper asymptotic behavior, violating basic thermodynamic principles. A few possible mitigation strategies are enforcing physical asymptotic behavior through architectural constraints using cutoff functions and a fallback to physics-based potentials in high-uncertainty regions \cite{Illarionov2023-ol}.

\subsection{The Curse of Dimensionality Persists:}
Despite architectural innovations, PIML still struggles with truly high-dimensional systems \cite{Cho2023-bf}. A 1,000-residue protein has $\sim$30,000 degrees of freedom; even with aggressive coarse-graining, the effective dimensionality remains high. PINNs in dimensions $>$20 often require prohibitive numbers of collocation points to adequately sample the residual loss. Operator networks need exponentially larger training sets to cover the input space. Gradient-based training becomes unstable due to the concentration of measure phenomenon. Current workarounds include dimensionality reduction via collective variables (VAMPnets, autoencoders), Hierarchical/multi-resolution architectures, custom Deep Operators and exploiting sparsity and locality in molecular interactions \cite{Chen2023-dw}. Despite this, the open problem of scaling PIML to large biomolecular assemblies with thousands to millions of atoms remains largely unsolved.

\subsection{Numerical Stiffness and Training Pathologies:}

Biomolecular systems exhibit extreme multi-timescale stiffness, bond vibrations at femtoseconds, conformational changes at microseconds, and folding at milliseconds. This creates severe training challenges. Standard neural networks preferentially learn low-frequency modes, systematically missing fast dynamics \cite{Wang2022-au}. This is well-documented: vanilla MLPs struggle to fit high-frequency functions even with ample capacity~\cite{Raissi2019-rz}. Choosing weighting coefficients $\lambda_{r}$, $\lambda_{b}$, $\lambda_{0}$ in PINN losses (Eq. \ref{pinn_full} Section~\ref{subsec:pinns}) remains an art. Poor choices lead to:
\begin{itemize}[leftmargin=1.25em,itemsep=2pt]
\item Overemphasis on data $\Rightarrow$ physics violations, poor generalization
\item Overemphasis on residuals $\Rightarrow$ ignoring available data, slow convergence
\item Imbalanced terms $\Rightarrow$ gradient pathologies, training collapse
\end{itemize}
Backpropagation through long MD rollouts ($>$10$^4$ steps) suffers from exponential growth or decay of gradients. Standard BPTT is infeasible for biologically relevant timescales.
A few mitigation strategies include:
\begin{itemize}[leftmargin=1.25em,itemsep=2pt]
\item \textbf{Adaptive loss weighting:} Automatically balancing loss terms using gradient statistics or neural tangent kernel theory \cite{Zhang2025-yn}
\item \textbf{Curriculum learning:} Training on easy physics first (equilibrium), gradually introducing complexity (dynamics, rare events) \cite{Bekele2024-tw, Bischof2025-nu}.
\item \textbf{Symplectic/Hamiltonian architectures:} Hard-wire energy conservation to reduce drift
\item \textbf{Multiple shooting:} Breaking long trajectories into short segments with consistent boundary conditions to ease computation.
\end{itemize}

\subsection{Nonequilibrium Generalization:}
Models trained on equilibrium data often fail catastrophically when applied to driven systems. The problem is subtle; equilibrium-trained potentials may conserve energy but violate more stringent constraints on entropy production or fluctuation theorems.

\textbf{Failure mode:} A learned force field that perfectly reproduces equilibrium free energies may:
\begin{itemize}[leftmargin=1.25em,itemsep=2pt]
\item Violate the Jarzynski equality: $\langle e^{-\beta W}\rangle \neq e^{-\beta \Delta F}$ \cite{Jarzynski1997-ug}
\item Predict negative entropy production: $\langle \Sigma \rangle < 0$ in a driven steady state
\item Violate Crooks' theorem: incorrect forward/reverse path probability ratios \cite{Crooks1999-xv}
\end{itemize}

Protein motors, membrane transporters, and enzymatic cycles operate far from equilibrium. Predicting their function requires respecting nonequilibrium thermodynamics.
Some possible solutions include:
\begin{itemize}[leftmargin=1.25em,itemsep=2pt]
\item Augmenting the training with nonequilibrium trajectories (pulling, flow, temperature ramps)
\item Adding path-integral loss terms enforcing Jarzynski/Crooks identities
\item Constraining the drift and diffusion to satisfy detailed balance at equilibrium, FDT coupling out of equilibrium \cite{Hu2024-kz}
\end{itemize}

\subsection{Data Quality and Bias}

\textbf{Garbage in, garbage out still applies;} PIML cannot fix fundamentally flawed data.

\textbf{Common data issues:}
\begin{itemize}[leftmargin=1.25em,itemsep=2pt]
\item \textbf{MD data inherits force field biases:} Training on AMBER-generated trajectories bakes in AMBER's systematic errors \cite{Zapletal2020-lr, Greener2024-nx}
\item \textbf{QM data has functional dependence:} DFT energies at the B3LYP level differ systematically from CCSD(T); ML models learn these biases \cite{Mardirossian2017-ix, Goerigk2017-pb}
\item \textbf{Experimental data is sparse and noisy:} X-ray structures have resolution limits; NMR observables are ensemble averages; cryo-EM maps are low resolution \cite{Cooper2011-vq, Renaud2018-tn}
\item \textbf{Sampling bias:} Stable states are overrepresented; transition states and rare events are undersampled \cite{Bernardi2015-js}
\end{itemize}

Bias in training data can propagate directly into physics-informed models, causing them to reproduce systematic errors rather than underlying physical truth, especially in sparsely sampled or skewed regions of chemical and conformational space. Such models often extrapolate poorly to unseen chemistries, environments, or boundary conditions, and their uncertainty estimates can become misleadingly overconfident in regimes where data are weakest. To mitigate these issues, best practices emphasize cross-validation against multiple independent data sources, including orthogonal experiments and higher-level theoretical calculations \cite{thuerey2021pbdl}. Active learning strategies can be used to target undersampled regions and improve coverage of the relevant physical landscape. Finally, benchmarking on genuinely out-of-sample test sets drawn from different sources provides a more reliable measure of true generalization performance and can reveal failure modes that remain hidden under standard random-split validation \cite{Vandermause2019-bo}.

\vspace{0.25em}

\begin{table}[H]
\centering
\small
\caption{Key limitations and practical mitigations in physics-informed biomolecular ML.}
\label{tab:limits_mitigations}
\renewcommand{\arraystretch}{1.12}
\begin{tabular}{@{}p{3.4cm}p{5.6cm}p{4.6cm}@{}}
\toprule
\textbf{Limitation} & \textbf{Symptom} & \textbf{Mitigation (refs)} \\ \midrule
Multiscale mismatch & Drift across AA$\leftrightarrow$CG; $\Delta F$ inconsistency & Free-energy/flux matching; hybrid closures with RE/VFM objectives \cite{Jia2020-dc} \\
Data sparsity \& bias & Overconfident errors; missed rare events & Bayesian/ensemble UQ; active learning for MFPT/$\Delta F$ reduction \cite{Vandermause2020-hz} \\
Stiff dynamics \& spectral bias & Unstable training; degraded long rollouts & Symplectic/HNN/LNN priors; curriculum, multiple shooting \cite{Raissi2019-rz,thuerey2021pbdl} \\
Nonequilibrium generalization & Violated detailed balance; wrong work stats & Pathwise losses enforcing Jarzynski/Crooks; reversible operators \cite{Karniadakis2021-hr} \\
Reproducibility \& compute & Sensitivity to seeds/backends; high memory & Fixed protocols; adjoint/checkpointing; report seeds/hardware \cite{thuerey2021pbdl} \\ 
\bottomrule
\end{tabular}
\end{table}

\subsection{When NOT to Use PIML}

Physics-informed ML is \textbf{not} a universal solution. Classical methods remain superior when:

\begin{enumerate}[leftmargin=1.5em,itemsep=2pt]
\item \textbf{System is low-dimensional with known physics:} Solving the 1D Schrödinger equation analytically beats any neural approximation
\item \textbf{High-quality transferable force fields exist:} AMBER/CHARMM for biomolecules in explicit water are mature, well-validated, and computationally efficient
\item \textbf{Data is extremely sparse ($<$100 samples):} Unless problem has strong structure, insufficient data for meaningful learning
\item \textbf{Problem is well-served by existing enhanced sampling:} Metadynamics or replica exchange may be simpler and more reliable than building a custom PIML pipeline
\end{enumerate}

\textbf{Decision framework:} PIML is most valuable when (a) physics is partially known but incomplete, (b) data is available but expensive, (c) many related problems must be solved, and (d) interpretability can be sacrificed for accuracy.

\section{Future Directions}\label{sec:future}
\noindent
The next decade will determine whether PIML becomes the default framework for molecular simulation or remains a specialized technique. Success depends on solving technical challenges while building community infrastructure and interdisciplinary expertise.

\textbf{Mechanistic Inductive Biases in Architectures:}
Current PIML often treats physics as a "soft" regularizer (loss term). The frontier is \emph{hard-wiring} physical structure into architecture, making violations impossible rather than merely penalized.
Concrete research directions include models that will \emph{embed} symmetries, conservation, and variational structure, curbing spurious drift and improving cross-chemistry generalization. In practice, this means SE(3)-equivariant message passing with explicit charge channels, symplectic updates for long horizon stability, and noether-style penalties that can tie invariances to conserved quantities. Examples include HNN \cite{Greydanus2019-hi}/LNN \cite{Cranmer2020-ej} layers for energy-consistent rollouts, equivariant force heads that satisfy conservative curl-free fields, and constraint-aware graph kernels that respect bond/angle manifolds. Such biases should transfer across solvents, protonation states, and metal coordination without re-tuning, reducing any “leaky” extrapolation.

\noindent
\textbf{Differentiable Simulation Ecosystems}
The ultimate vision is a  gradient flow from \emph{experimental observable} $\rightarrow$ learned representation $\rightarrow$ physics simulator $\rightarrow$ predicted measurement $\rightarrow$ experimental design. GPU-native, end-to-end differentiable MD will complete the loop by simultaneously optimizing integrators, thermostats, and ML potentials directly against experiment-level objectives, blending PINN/operator residual constraints with trajectory-level losses to obtain consistently calibrated dynamics. Concretely, differentiable Verlet schemes, barostats, and constraint solvers wired to losses from SAXS profiles, FRET efficiencies, cryo-EM densities, and NMR couplings, plus operator residuals for Smoluchowski/Fokker-Planck consistency, are expected to be studied. Multi-fidelity training will fuse cheap CG trajectories with short all-atom bursts and sparse experiments through differentiable reweighting and control variates. 
The fundamental challenges arise from the discreteness of conformational space, which necessitates relaxation or reinforcement-learning strategies, combined with the high computational cost of long-horizon rollouts and the presence of multiple, often conflicting objectives that require principled Pareto-optimal trade-offs.

\noindent
\textbf{Physics-Aware Generative Models}
The idea is that Flow/diffusion models regularized by Boltzmann and pathwise constraints will produce reweightable ensembles and transition paths, enabling accelerated yet thermodynamically admissible sampling. Energy-guided diffusion and Boltzmann generators can target rare states (e.g., cryptic pockets) while preserving detailed balance via path-action penalties. Post-hoc reweighting engine across temperatures, ionic strengths, or ligand states could turn a single generator into a family of thermodynamic ensembles.

\noindent
\textbf{Uncertainty-Driven Active Learning}
Slowly we see a trend of using Hybrid-Bayesian PIML that propagates epistemic and aleatoric uncertainty through simulators to guide adaptive sampling and experiment design toward maximal information gain on free energies and rates \cite{Yang2021-hy}. This might mitigate the current overconfident PIML models that adhere to non-conformity scoring based on violations. Practical tools include ensemble/Laplace posteriors over force fields, GP heads on CVs, and Bayesian PINNs with physics-aware likelihoods. Acquisition functions that target \emph{physics}, maximizing variance reduction in $\Delta F$, MFPTs, rather than generic loss. This enables on-the-fly decisions, such as where to seed enhanced sampling or which mutational scans most reduce uncertainty in binding kinetics and provide a certified error bound for long-time predictions.

\noindent
\textbf{Physics-Grounded Foundation Models}
Inspired by large language models, several groups are building "molecular foundation models" pretrained on massive simulation and experimental datasets.
\textbf{Key open questions:}
\begin{enumerate}[leftmargin=1.5em,itemsep=2pt]
\item \textbf{Tokenization:} How to embed continuous molecular configurations into discrete tokens? Voxelization? Graph encoding? Point clouds?
\item \textbf{Self-supervised objective:} What is the "next-token prediction" for molecules? Predicting masked atoms? Forecasting dynamics? Denoising structures?
\item \textbf{Multi-modal integration:} Can a single model handle small molecules, proteins, nucleic acids, and materials at multiple resolutions (QM, AA, CG)?
\item \textbf{Transfer learning:} How much fine-tuning is needed? Can we do "few-shot" force field learning with $<$100 examples?
\end{enumerate}
Approaches to pretraining over sequence/structure corpora with symmetry- and physics-aware objectives will yield universal priors whose embeddings expose gradients, energetics, and slow modes as queryable operators. Beyond masked-token tasks, objectives can include equivariant force/energy prediction, contrastive views between MD fragments and experimental densities, and curriculum tasks that tie local chemistry to mesoscopic dynamics\cite{Yuan2025-pd}. Lightweight adapters that can specialize these priors to new chemistries or environments, while probes that can retrieve CVs, committors, or elastic response without full fine-tuning \cite{Kalfon2025-yr}. Such models promise a single backbone that supports design, simulation warm-starts, and kinetics analysis with minimal data.

\section{Conclusion}\label{sec:conclusion}

Machine learning devoid of physics is prone to overfitting, inadequate extrapolation, and misinterpretation of nature's guiding symmetries; conversely, physics lacking machine learning frequently does not scale, undersamples, and inadequately fits the intricacies of actual data. The next era of biomolecular modeling will arise not from a singular source but from their integration. Physics-informed and physics-based learning frameworks, including PINNs, neural operators, differentiable simulators, and probabilistic surrogates, are starting to close this gap. These hybrid methodologies reframe conventional numerical simulation as an optimization framework in which physical consistency, differentiability, and uncertainty are learned concurrently. By incorporating conservation rules, symmetries, and thermodynamic restrictions directly into neural networks or loss functions, these models become not only predictive but also interpretable and transferable across many scales. 

Looking ahead, three directions define the frontier. First, the development of \emph{differentiable biomolecular simulators}, integrating automatic differentiation with MD engines, will enable end-to-end optimization of molecular potentials and experimental fits. Second, \emph{probabilistic physics-informed learning} will quantify uncertainty and ensemble variability across dynamic landscapes. Third, \emph{hybrid generative-mechanistic models} will move beyond reproducing trajectories to discovering emergent coordinates and physical laws. Ultimately, the goal of physics-informed machine learning in biomolecular dynamics is not to replace simulation but to elevate it to learn “\emph{with}” the laws of physics, not “\emph{in spite of}” them. By enforcing physical truth while embracing data-driven flexibility, these methods mark the beginning of a new paradigm: hybrid, interpretable, mechanistic, and differentiable molecular science.

\section*{Acknowledgments}

The author acknowledges Prof. Shina Caroline Lynn Kamerlin (School of Chemistry \& Biochemistry, Georgia Institute of Technology) for valuable discussions and feedback that substantially improved the clarity and scope of this review.

\printbibliography

\clearpage
\appendix
\section{Abbreviations}\label{app:abbrev}

\begin{multicols}{2}
\raggedcolumns
\begin{itemize}[leftmargin=1.1em,itemsep=2pt,topsep=2pt]
  \item \textbf{PIML}: Physics-Informed Machine Learning
  \item \textbf{PINN}: Physics-Informed Neural Network
  \item \textbf{SDE-PINN}: Stochastic Differential Equation PINN
  \item \textbf{HNN/LNN}: Hamiltonian/Lagrangian Neural Network
  \item \textbf{MD/AA/CG}: Molecular Dynamics / All-Atom / Coarse-Grained
  \item \textbf{FES/CV}: Free-Energy Surface / Collective Variable
  \item \textbf{UQ}: Uncertainty Quantification
  \item \textbf{PDE}: Partial Differential Equation
  \item \textbf{GNN}: Graph Neural Network
  \item \textbf{E(3)/SE(3)}: Euclidean/Special Euclidean symmetry in 3D
  \item \textbf{QM / QM/MM}: Quantum Mechanics / Quantum Mechanics–Molecular Mechanics
  \item \textbf{JAX-MD}: Differentiable MD library in JAX
  \item \textbf{TorchMD / TorchMD-Net}: Differentiable MD / equivariant ML potential
  \item \textbf{DeepMD}: Deep Potential Molecular Dynamics
  \item \textbf{MLIP(s)}: Machine-Learned Interatomic Potential(s)
  \item \textbf{FDT}: Fluctuation–Dissipation Theorem
  \item \textbf{NESS}: Nonequilibrium Steady State
  \item \textbf{VAMP / VAMPnets}: Variational Approach for Markov Processes / neural impl.
  \item \textbf{TICA}: Time-lagged Independent Component Analysis
  \item \textbf{PMF}: Potential of Mean Force
  \item \textbf{GP / GPR}: Gaussian Process / Gaussian Process Regression
  \item \textbf{KL}: Kullback–Leibler (divergence)
  \item \textbf{MFPT}: Mean First-Passage Time
  \item \textbf{NVE/NVT/NPT}: Energy/Temp/Pressure ensembles
  \item \textbf{BPTT}: Backpropagation Through Time
  \item \textbf{AD}: Automatic Differentiation
  \item \textbf{SAXS / FRET / cryo-EM / NMR}: Scattering / Transfer / EM / Resonance
  \item \textbf{RDC / NOE}: Residual Dipolar Coupling / Nuclear Overhauser Effect
  \item \textbf{RDF / $S(q)$}: Radial Distribution Function / Static Structure Factor
  \item \textbf{VACF}: Velocity Autocorrelation Function
  \item \textbf{IC/BC}: Initial Condition / Boundary Condition
  \item \textbf{PLM}: Protein Language Model
  \item \textbf{FM / RE}: Force Matching / Relative Entropy
\end{itemize}
\end{multicols}

\end{document}